\documentclass[aps,prl,nofootinbib,amsfonts,tightenlines,twocolumn,superscriptaddress]{revtex4-1}

\usepackage{graphicx} 
\usepackage{amsmath}
\usepackage{amssymb}
\usepackage{ae} 
\usepackage{aecompl}
\usepackage{bm} 
\usepackage[usenames,dvipsnames]{color}
\usepackage[pdfa,bookmarks=true,bookmarksnumbered=true]{hyperref}
 \hypersetup{
     colorlinks=true,
     linkcolor=Blue,          % color of internal links (change box color 
     citecolor=Mahogany,
     urlcolor=Mahogany
     }

\usepackage{amsmath}
\usepackage{amssymb}
\usepackage{epsfig}
\usepackage{fontenc}
\usepackage{graphicx}
\usepackage[lofdepth,lotdepth,caption=false]{subfig}
\usepackage{color}
\usepackage{booktabs}
\usepackage{tabularx}
\usepackage{multirow}
\usepackage{longtable}
\usepackage{dcolumn}
\usepackage{rotating}%

\newcommand{\bi}{\begin{itemize}}
\newcommand{\ei}{\end{itemize}}

\def\beq{\begin{equation}}
\def\eeq{\end{equation}}

\newcommand{\bea}{\begin{eqnarray}}
\newcommand{\eea}{\end{eqnarray}}

\newcommand{\eee}{\varepsilon_{ee}}

\newcommand{\eeta}{|\varepsilon_{e\tau}|}

\newcommand{\eema}{|\varepsilon_{e\mu}|}

\newcommand{\eetp}{\varphi_{e\tau}}

\newcommand{\eemp}{\varphi_{e\mu}}
\def\epsilon{\varepsilon}

\newcommand\sch{Schr$\ddot{\rm o}$dinger~}

\def\<{\langle}
\def\>{\rangle}

\def\lsim{\mathrel{\rlap{\lower4pt\hbox{\hskip1pt$\sim$}}
    \raise1pt\hbox{$<$}}}         %less than or approx. symbol
\def\gsim{\mathrel{\rlap{\lower4pt\hbox{\hskip1pt$\sim$}}
    \raise1pt\hbox{$>$}}}         %greater than or approx. symbol

\begin{document}

\title{Extricating New Physics Scenarios at DUNE with High Energy Beams}
\author{Mehedi Masud}
%\email{masud@ific.uv.es}
\affiliation{Astroparticle and High Energy Physics Group, Institut de F\'{i}sica Corpuscular --
  C.S.I.C./Universitat de Val\`{e}ncia, Parc Cientific de Paterna. 
  C/Catedratico Jos\'e Beltr\'an, 2 E-46980 Paterna (Val\`{e}ncia) - Spain}
\affiliation{Harish-Chandra Research Institute, Chattnag Road, Allahabad 211 019, India}
    \author{Mary Bishai}
 % \email{mbishai@bnl.gov}
\affiliation{Brookhaven National Laboratory,  P.O. Box 5000, Upton, NY 11973, 
USA}
\author{Poonam Mehta}
%\email{pm@jnu.ac.in}
\affiliation{School of Physical Sciences, Jawaharlal Nehru University, 
      New Delhi 110067, India}
\date{\today}

\begin{abstract}
 The proposed Deep Underground Neutrino Experiment (DUNE) utilizes a wide-band on-axis tunable
 muon-(anti)neutrino beam with a baseline of 1300km to search for CP violation with high precision. 
 Given the long baseline, DUNE is also sensitive to effects due to non-standard
 neutrino interactions (NSI) which can interfere with the standard 3-flavor
 oscillation paradigm.
In this Letter, we exploit the tunability of the DUNE neutrino beam
over a wide-range of energies and utilize a new theoretical metric to devise an experimental strategy
for separating oscillation effects due to NSI from the standard 3-flavor oscillation scenario. 
Using our metric, we obtain an optimal combination of beam tunes and
distribution of run times in neutrino and anti-neutrino modes that
would enable DUNE to isolate new physics scenarios from the standard.
To the best of our knowledge, our strategy is entirely new and has not
been reported elsewhere.
\end{abstract}

\maketitle

\section{Introduction}
 % {\underline{\bf{Introduction :-}}}
Neutrino oscillations among the three flavours have been firmly
established and the experimental confirmation of neutrino oscillations
vindicates that the Standard Model (SM) of particle physics is
incomplete~\cite{nobel2015}.  The minimal extension of SM invokes a
mechanism to generate tiny neutrino masses while retaining the
interactions as predicted in the SM. We refer to this minimal model as
Standard Interactions (SI).

Most of the parameters responsible for standard 3-flavor neutrino
oscillations have been measured with fairly good precision except for
a few~\cite{Bergstrom:2015rba}.  Some of the yet unresolved questions
in neutrino physics include whether CP is violated, if the neutrino
mass hierarchy is normal or inverted and what the correct octant of
$\theta_{23}$ is.  Detecting CP violation is one of the most
challenging goals in particle physics. An attractive possibility to
measure the neutrino CP phase is via long-baseline accelerator
experiments such as the Deep Underground Neutrino Experiment
(DUNE)~\cite{Bass:2013vcg}. DUNE and the facility that
will support it, the Long-Baseline Neutrino Facility (LBNF), will be an internationally
designed, coordinated and funded program, hosted at the Fermi National Accelerator
Laboratory (Fermilab) in Batavia, Illinois~\cite{Acciarri:2015uup}.

In the presence of new physics effects, clean extraction of the CP violating
phase becomes a formidable
task~\cite{Nunokawa:2007qh,Rout:2017udo}. In fact, a given measured
value of CP phase could be a hint of new
physics~\cite{Forero:2016cmb,Miranda:2016wdr}. In our earlier work,
we have pointed out that there are degeneracies within the large
parameter space in the presence of non-standard interactions
(NSI)~\cite{Ohlsson:2012kf,Masud:2015xva,deGouvea:2015ndi,Coloma:2015kiu,Liao:2016hsa,
  Masud:2016bvp,Masud:2016gcl} and the need to devise ways to
distinguish between the standard paradigm and new physics scenarios has
been extensively discussed (for new physics scenarios with extra
sterile neutrinos
see~\cite{Gandhi:2015xza,Dutta:2016glq,Agarwalla:2016xlg} and for
non-unitarity see~\cite{Dutta:2016vcc,Escrihuela:2016ube}).

In a novel approach, we use experimental handles that could prove
useful to differentiate between the standard scenario (with only one
source of CP violation) and new physics scenarios (which inevitably
bring in more parameters including new sources of CP violating
phases).
Recent studies have explored the sensitivities to SI paramaters and
the synergies between experiments (DUNE and T2HK) using different
baselines and neutrino beam energies~\cite{Ballett:2016daj}. In this
study, for the first time, we explore sensitivities to both SI and NSI
effects at a fixed baseline over a large range of $L/E$ using DUNE's
unique broad-band tunable beam. Different wide-band fluxes can be
experimentally achieved using the DUNE NuMI-style reference beam
design~\cite{Acciarri:2015uup} by simply varying the target and horn
placement~\cite{Adamson:2015dkw}. We propose a new theoretical metric
that allows us to optimize experimental strategies and beam tunes for
clean inference of the leptonic CP phase in the presence of new
physics.

\section{Non-Standard $\nu$ Interaction Model}
%  {\underline{\bf{Non-Standard $\nu$ Interaction Model :-}}}
 The effective Hamiltonian in the flavour basis entering the 
 \sch equation for neutrino propagation is given by
 \begin{eqnarray}
 \label{hexpand} 
 {\mathcal
H}^{}_{\mathrm{f}} &=&   {\mathcal
H}^{}_{\mathrm{v} } +  {\mathcal
H}^{}_{\mathrm{SI} } +  {\mathcal
H}^{}_{\mathrm{NSI}} 
\nonumber 
\\
&
=&\lambda \Bigg\{ {\mathcal U} \left(
\begin{array}{ccc}
0   &  &  \\  &  r_\lambda &   \\ 
 &  & 1 \\
\end{array} 
\right) {\mathcal U}^\dagger  + r_A   \left(
\begin{array}{ccc}
1  & 0 & 0 \\
0 &  0 & 0  \\ 
0 & 0 & 0 \\ 
\end{array} 
\right)  \nonumber\\
&+&
 {r_A}   \left(
\begin{array}{ccc}
\epsilon_{ee}  & \epsilon_{e \mu}  & 
\epsilon_{e \tau}  \\ {\epsilon_{e\mu} }^ \star & 
\epsilon_{\mu \mu} &   \epsilon_{\mu \tau} \\ 
{\epsilon_{e \tau}}^\star & {\epsilon_{\mu \tau}}^\star 
& \epsilon_{\tau \tau}\\
\end{array} 
\right) \Bigg\}  \ ,
 \end{eqnarray} 
where 
\begin{equation}
\lambda \equiv \frac{\delta m^2_{31}}{2 E}  \quad ; \quad
r_{\lambda} \equiv \frac{\delta m^2_{21}}{\delta m^2_{31}} \quad  ; 
\quad  r_{A} \equiv \frac{A (x)}{\delta m^2_{31}} \ .
\label{dimless}
\end{equation}
and  
 $A (x)= 2 \sqrt{2}  	E G_F n_e (x)$  where  $n_e$ is the electron
number density. 
The three terms in Eq.~\ref{hexpand} are due to vacuum, matter with
standard interaction (SI) and matter with NSI respectively. For the
NSI case, the ${\varepsilon}_{\alpha \beta} \, (\equiv |\varepsilon
_{\alpha \beta}| \, e^{i \varphi_{\alpha\beta}})$ are complex
parameters which appear in ${\cal H}_{NSI}$.  As a result of the
hermiticity of the Hamiltonian, we have nine additional parameters
(three phases and six amplitudes appearing ${\cal H}_{NSI}$).
Thus, there are new genuine sources of CP violation as well as new
fake sources of CP violation (aka matter effects) that can change the
asymmetries even further. For more details,
see~\cite{Masud:2015xva,Masud:2016bvp,Masud:2016gcl} and references
therein.

To quantify the separation of physics scenarios (SI-NSI separation),
we define\footnote{The definition of the $\chi^{2}$ in
  Eq.~\ref{eq:chisq_si_nsi} includes only statistical effects and
  facilitates our understanding. The systematic effects are taken into
  account in the numerical results.}  the (statistical) $\chi^2$ as
follows in order to interpret results -
 \begin{eqnarray}
 \label{eq:chisq_si_nsi}
 \chi^{2}(\delta_{tr}) & = &
  \min_{\delta_{ts}}  \sum_{i=1}^{x}  \sum_{j}^{2} 
 \nonumber\\
 && \!\!\!\!\!\!\!\!\!
  \frac{\bigg[N_{NSI}^{i,j}(\delta_{tr},|\varepsilon|,\varphi) - N_{SI}^{i,j} (\delta_{ts}\in[-\pi,\pi])\bigg]^{2}}{N_{NSI}^{i,j} (\delta_{tr},|\varepsilon|,\varphi)}
\end{eqnarray}
where, we have marginalised over the standard CP phase $\delta$ in the
test dataset.  This $\chi^{2}$ was calculated using a set of
conservative values of the non-zero NSI parameters ($\eema$ = 0.04,
$\eeta$ = 0.04\ $\eee$ = 0.4)~\cite{Biggio:2009nt,Davidson:2003ha}.

\section{Neutrino Beam Tunes}  
%{\underline{\bf{Neutrino Beam Tunes :-}}}

\begin{figure}[tb!]
\centering
\includegraphics[width=0.5\textwidth]
{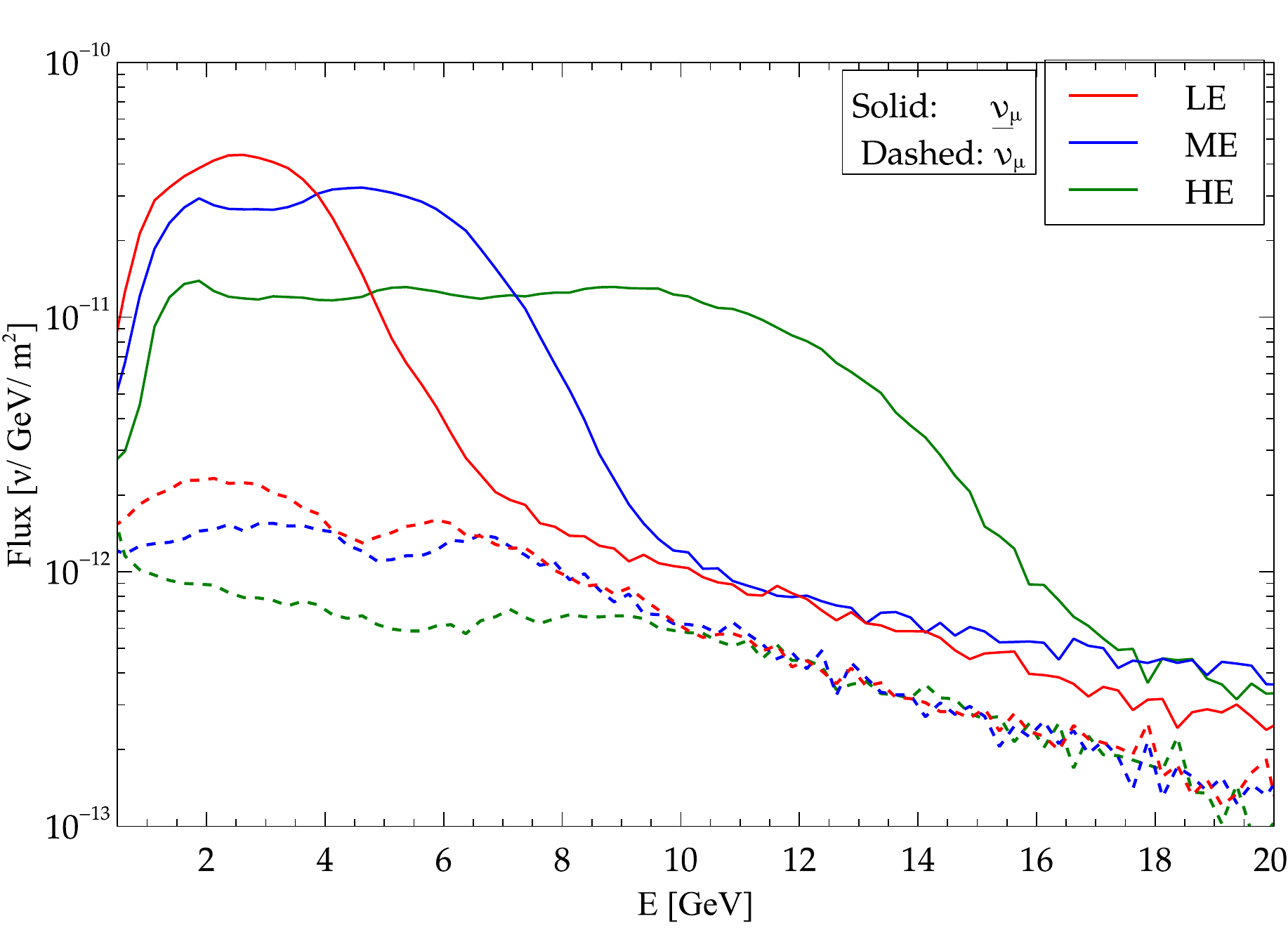}
\caption{\footnotesize{Comparison of the different flux tunes (LE, ME, HE) in the neutrino running mode. 
}}
\label{fig:1}
\end{figure}

For this study, we considered three wide-band beam tunes obtained from
a full Geant4 simulation~\cite{Agostinelli:2002hh,Allison:2006ve} of a neutrino beamline using NuMI-style
focusing. The tunes considered are: low energy (LE); medium energy
(ME); and high energy (HE) as shown in Fig.~\ref{fig:1}. These tunes
are consistent with what could be achieved by the LBNF facility. The
energy range considered is $E=0.5-20$ GeV. The beamline parameters
assumed for the different design fluxes used in our sensitivity
calculations are given in Table~\ref{tab:cdr}
(see~\cite{2013arXiv1307.7335L,Alion:2016uaj}).

%%%%%%%%%%%%%% 
%%%%%%%%%%%%%%%% 

\begin{table}[htb!]
\centering
{\scriptsize{
\begin{tabular}{| l | l l l|}
\hline
%&&&\\
Parameter & LE & ME & HE \\
%&&&\\
\hline \hline
%&&&\\
Proton Beam  & \multicolumn{3}{c|}{$E_{p^+} =120$ GeV, 1.2 - 2.4 MW } \\
Focusing &  \multicolumn{3}{c|}{2 NuMI horns, 230kA, 6.6 m apart }\\
Target location & -25cm & -1.0m & -2.5m \\
Decay pipe length & 250 m & 250 m  & 250 m \\
Decay pipe diameter & 4 m & 4 m & 4m  \\
%&&&\\
\hline
\end{tabular}
}}
\caption{\label{tab:cdr} 
Beamline parameters assumed for the different design fluxes used 
in our sensitivity calculations~\cite{2013arXiv1307.7335L,Alion:2016uaj}. The 
 target is a thin Be cylinder 2 interaction lengths long. The target location is given with respect to the upstream face of Horn 1.} 
\end{table}

\section{Results and Discussion}
%\underline{\bf{Results and discussion :-}}}

%-------------------------------------------------------

\begin{figure}[t!]
% Plots for the CP=0 case in the folder
\includegraphics[width=0.99\columnwidth]{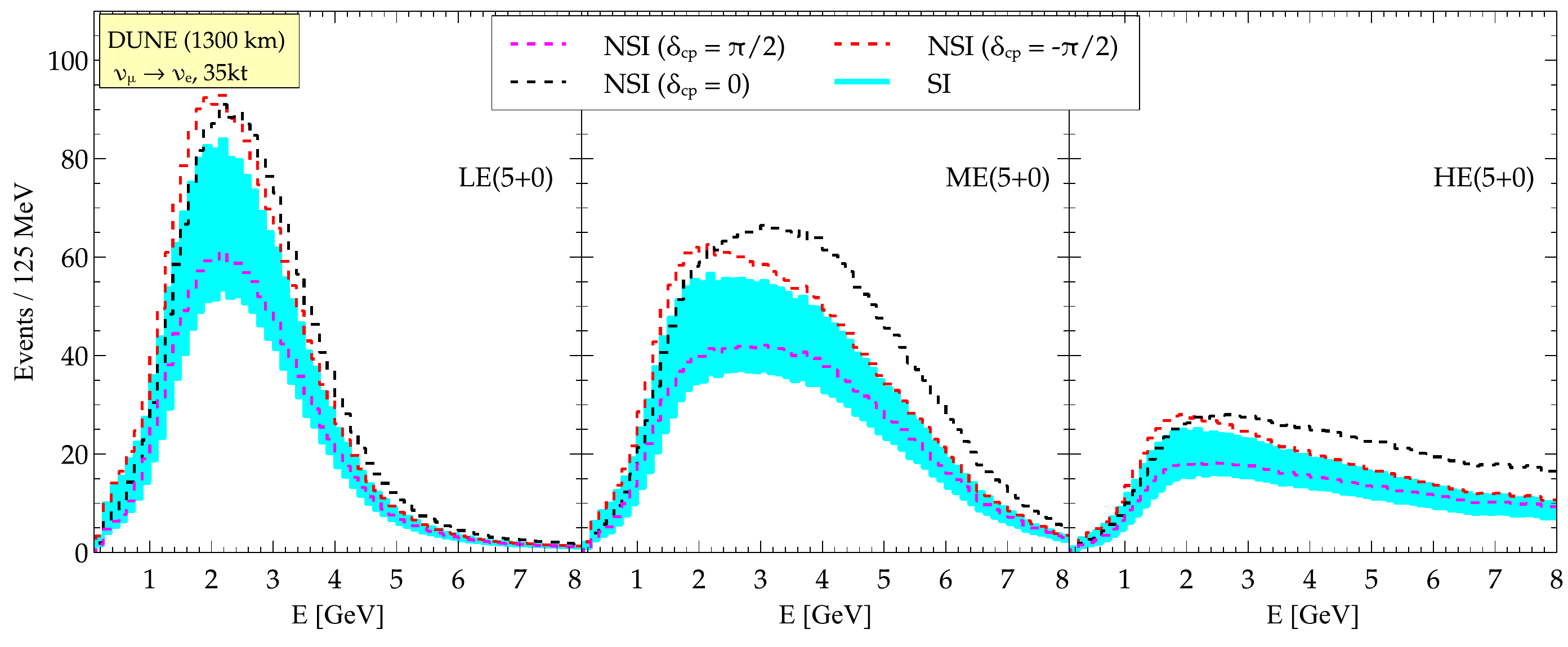}%
 \caption{
 Separation between SI $\nu_\mu \to \nu_e$ events (cyan band, red and magenta dashed 
lines) and NSI events (black dashed lines) at DUNE with LE (5+5), ME(5+5) and HE(5+5) beam tunes. The
black dashed line is for a CP conserving NSI scenario (all phases are set to zero). The cyan band corresponds to the SI case with the full
 variation of $\delta$. The other dashed lines are for NSI with different true values of
  $\delta$.
}
\label{fig:event_me_3beams}
\end{figure}

We have implemented a GLoBES \cite{Huber:2004ka,Huber:2007ji}
simulation of a $1300\,\text{km}$ baseline neutrino beam experiment
using a parameterization of the DUNE far detector response as
described in ~\cite{Alion:2016uaj}. We assume normal hierarchy (NH) in
all the plots.
 We show the variation in the $\nu_e$ event spectrum in
 Fig.~\ref{fig:event_me_3beams} for the LE, ME and HE beam tunes under
 SI-only and SI+NSI scenarios. In all beams, the red and magenta
 dashed lines (for $\delta \sim \pm\pi/2$ with NSI) lie almost
 completely within the cyan band (SI for $\delta \in [-\pi,\pi]$) and that makes the separation
 between the two considered scenarios more difficult.  
 The black dashed lines (for $\delta \sim 0$ with NSI) lie farthest
 apart from the cyan band (SI). This particular feature results in
 better separability between the two considered scenarios at values of
 $\delta \sim 0$ (or $\pm \pi$). In addition, better
 separation is obtained at higher energies in the ME and HE beams.

\begin{figure}[t!]
% Plots for the CP=0 case in the folder
\includegraphics[width=0.99\columnwidth]{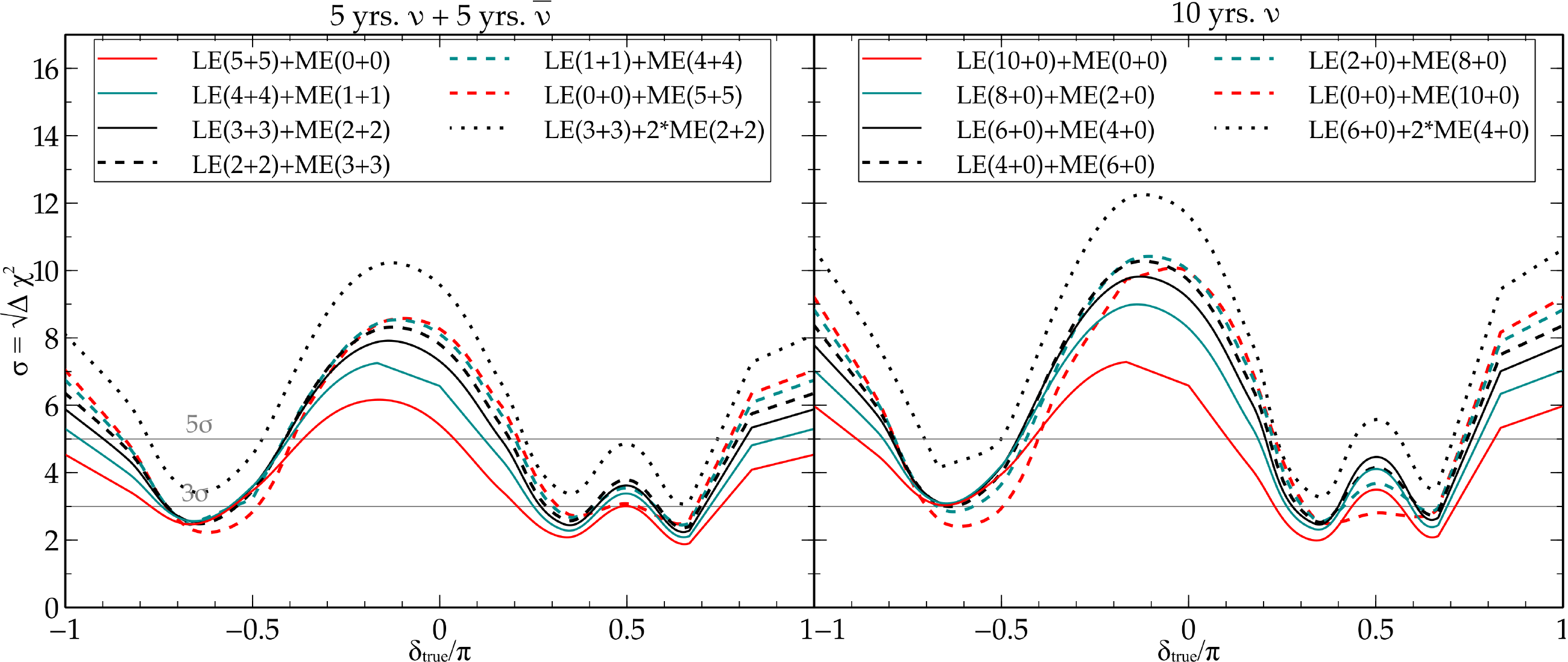}
 \caption{Separation between SI and NSI events at DUNE with different
   beam tunes at $\chi^2$ level. A CP conserving NSI scenario is
   assumed. The left column shows 5 years of neutrino and 5 years of
   anti-neutrino run times, while the right column depicts the case of 10
   years of neutrino run time only.}
\label{fig:opt_le_me}
\end{figure}

  In Fig.~\ref{fig:opt_le_me}, we show the ability of DUNE to separate
  SI from NSI using different combinations of beam tunes and running
  times at the $\chi^2$ level. The first panel is for an equal
  distribution of run time among neutrino and anti-neutrino modes
  while the second panel corresponds to running in neutrino-only mode
  with the same total run time.  For this analysis, we consider
  the energy range $E \in 0.5 - 20 $ GeV. A CP conserving NSI scenario
  is assumed in this plot\footnote{We assume $\eemp = \eetp = 0$.}. We
  have considered a combination of appearance ($\nu_\mu\to\nu_e$) and
  disappearance ($\nu_\mu\to\nu_\mu$) channels.  The solid and dashed
  lines assume a beam power of 1.2 MW for both LE and ME.  {{The
      dotted black line corresponds to an ME option upgraded to 2.4 MW
      which is planned for later stages of DUNE. }} We note that the
  dominant channel contributing to the distinction of different physics
  scenarios is the $\nu_\mu\to\nu_e$ channel irrespective of our choice 
  of beam tune. The $\nu_\mu \to \nu_\mu$
  channel adds somewhat to the total sensitivity but the $\nu_\mu\to\nu_\tau$
  contribution is negligible.
 
  As noted earlier, we find that the ability to separate between the
  two scenarios tends to increase at CP conserving values of $\delta$
  i.e. $\delta \sim 0, \pm \pi$. The dips seen near $\delta \sim \pm
  \pi/2$ (true) in Fig.~\ref{fig:opt_le_me} for all the LE and ME
  options imply the inability of DUNE to distinguish between the
  scenarios at those values.  If the standard beam power is assumed,
  one gets the best optimization for LE (2+2) + ME(3+3) (dashed black
  line). In fact, in general the different beam tunes and run time
  combinations other than LE only (solid red line) yield better
  results.  From the right panel, we can see that the best sensitivity
  at $\delta \sim 0$ is reached for LE (4+0) + ME (6+0) (dashed black
  line).  {{An upgrade of beam power in ME (dotted black line) to 2.4
      MW significantly improves the outcome.}}

\begin{figure}[t!]
%% Plots with only lines in the folder... also bi-prob plot for DUNE
%\includegraphics[width=0.95\columnwidth]{oscillogram_beam_comb_phi_em.pdf} %
\includegraphics[width=0.99\columnwidth]{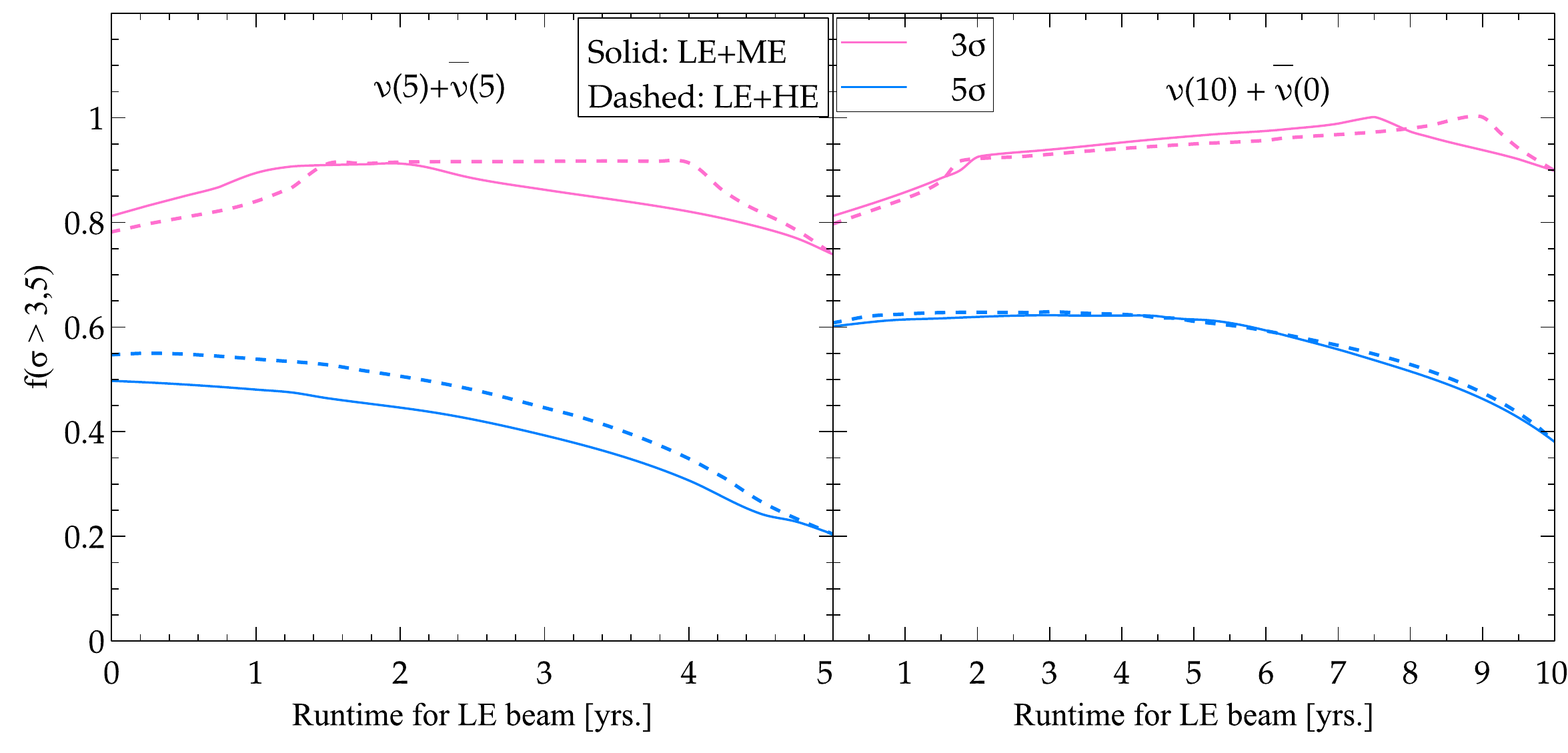} %
 \caption{The fraction of the values of $\delta$ for which SI and NSI
   scenarios can be distinguished above $3\sigma$ (magenta) and
   $5\sigma$ (blue) using different combinations of beam tunes.  }
\label{fig:fraction}
\end{figure}

Another important factor driving the sensitivity to SI-NSI separation
is the fraction of values of CP phase for which the sensitivity is
more than $3\sigma$ or $5\sigma$.  This quantity is plotted in
Fig.~\ref{fig:fraction} 
for the fraction lying above $3\sigma$ (magenta) and $5\sigma$ (blue)
as a function of the run time for a combination of LE and ME (or HE)
tuned beams. Both the panels are for a total run time of 10 years : the
left one showing the case of 5 years of $\nu$ and 5 years of
$\bar{\nu}$ run time while the right panel depicting the scenario of 10
years of $\nu$ run time alone.  In the left panel, the 5+5 years of
run time are distributed among the LE and ME (HE) beams for the solid
(dashed) lines in the following manner: $(x+x)$ years of LE beam $+
\big((5-x) + (5-x)\big)$ years of ME or HE  runtime  with
the run-time in one mode, $x$, in years along the x-axis. Similarly, the
runtime in the right panel has been distributed as $(x+0)$ years of LE
beam $+ \big((10-x) + 0\big)$ years of ME or HE run time. We note
as long as the distribution of runtime among the two different beams
is evenly distributed (approximately) in the LE and ME or HE
options, the fraction is close to its maximum value and remains almost
flat. However as we go to the extreme cases (ME or LE taken in
isolation for e.g.), the fraction sees a drop (this is more significant for LE
only towards the right edge).  So, this observation implies that one
must consider ME (HE) or both LE and ME (HE) beam tunes in order to
obtain the discrimination of new physics from standard at some given
level of significance.  We wish to stress that the fraction curves in
Fig.~\ref{fig:fraction} only show what portion of the sensitivity
curve lies above $3\sigma$ (or $5\sigma$), and not necessarily the
absolute value of the sensitivities. The estimate of the fraction of $\delta$ values
thus depends on the points of intersection of the sensitivity
curve with the $3\sigma$ (or $5\sigma$) horizontal lines in
Fig.~\ref{fig:opt_le_me}.

\begin{figure}[t!]
\includegraphics[width=0.99\columnwidth]{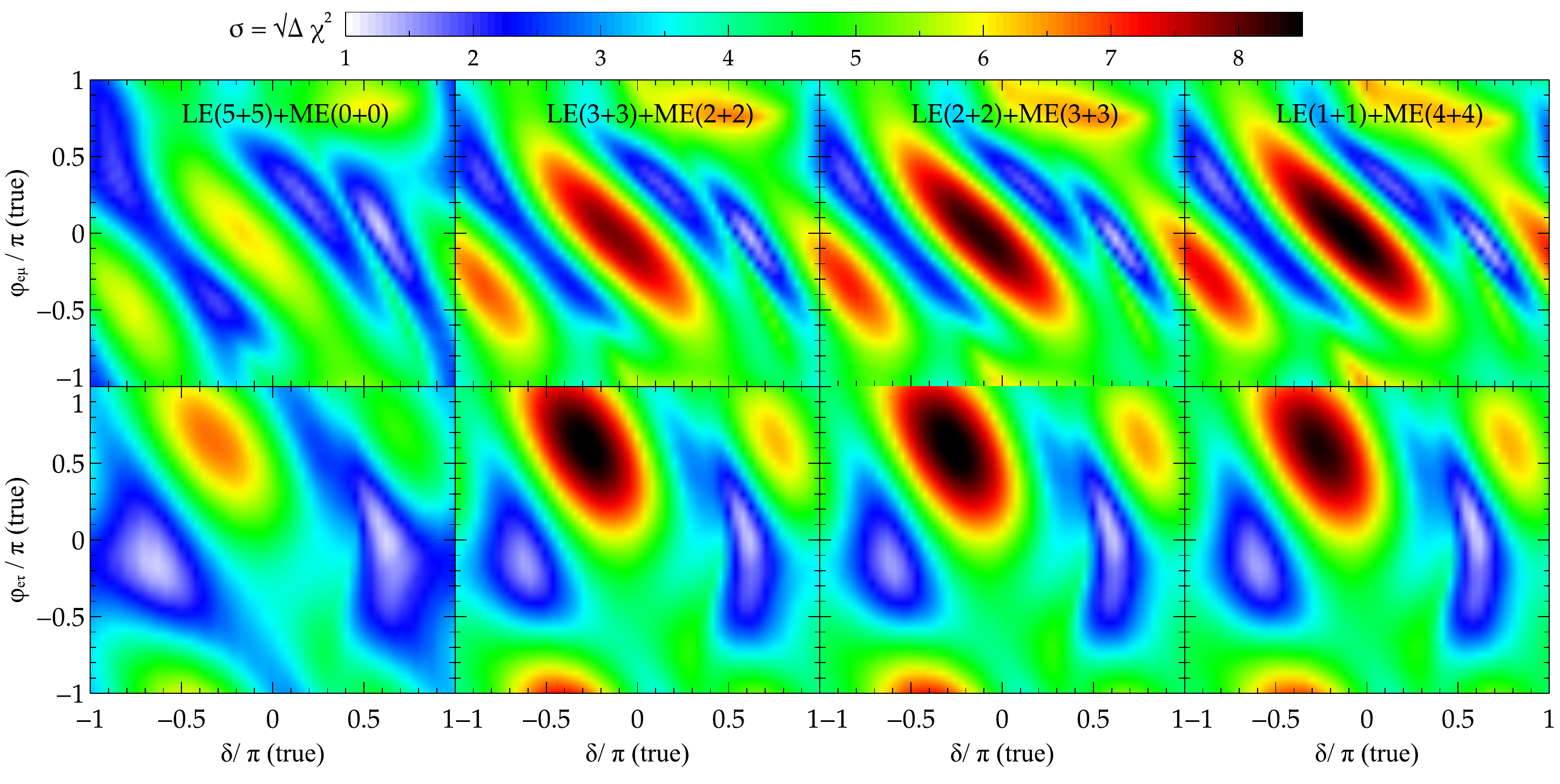} %
 \caption{Effect of non-zero NSI phases $\varepsilon_{e\mu}$ (top) and $\varepsilon_{e\tau}$ (bottom) on the ability to distinguish between SI and NSI. 
}
\label{fig:osc_nonzero_em}\end{figure}

In Fig.~\ref{fig:osc_nonzero_em}, we go beyond the CP conserving NSI
scenario considered so far and generalize Fig.~\ref{fig:opt_le_me} by
considering non-zero NSI phases. We show the ability to discriminate
between SI-NSI using oscillograms. The projection of the $\chi^2$  values
at $\eemp = 0$ in Fig.~\ref{fig:osc_nonzero_em} corresponds to
Fig.~\ref{fig:opt_le_me}.  For CP violating NSI scenarios, the peak
position in the sensitivity plot shifts left or right (with respect to
$\delta_{tr} = 0$) as shown in Fig.~\ref{fig:osc_nonzero_em}. The separation between scenarios
in Fig.~\ref{fig:osc_nonzero_em} is consistent with
Fig.~\ref{fig:opt_le_me} as the darker patches  are around $\delta \sim
0, \pm \pi$ while the lighter patches are around $\delta \sim \pm
\pi/2$.

%-------------------------------------------------------

%experimental set up

%-------------------------------------------------------

%%%%%%%%%%%%%%%%%%%%%%%%%%%%%%%%%%%%%%%%%%%%%%%%%%%%%%%%%%%%%%%%%%%

%{\underline{\bf{Summary :-}}}
\section{Summary}

It is crucial to separate different physics scenarios at DUNE to be
able to infer the parameters cleanly.  The key point of the study
reported in this Letter is to demonstrate the feasibility of using
DUNE's experimental flexibility to expand the physics reach beyond the
neutrino standard model. We have demonstrated that it is plausible to
have better separation of SI from NSI if we consider different
combinations of beam tunes and run times.  For the CP conserving NSI
scenario, a mix of LE (4+0) + ME (6+0) is close to optimal with
neutrino only mode. The results also show that LE+HE beam combinations give
slightly better results than LE+ME combinations with more LE than HE
in the mix. For the CP violating NSI scenario, the peak position in
the sensitivity plot shifts left or right (with respect to
$\delta_{tr} = 0$) as can be seen from Fig.~\ref{fig:osc_nonzero_em}.
We are currently expanding the study to include more new physics
scenarios.

\begin{acknowledgements}
It is a pleasure to thank Raj Gandhi for useful discussions and
critical comments on the manuscript.  This material is based upon work
supported by the Spanish grants FPA2014- 58183-P, SEV-2014-0398
(MINECO) and PROMETEOII/ 2014/084 (Generalitat Valenciana); the
U.S. Department of Energy, Office of Science, Office of High Energy
Physics under contract number DE-SC0012704; the Indian funding from
University Grants Commission under the second phase of University with
Potential of Excellence (UPE II) and DST-PURSE at JNU; and the
European Union's Horizon 2020 research and innovation programme under
Marie Sklodowska-Curie grant No 674896.
PM acknowledges support and kind hospitality from the particle physics group at
 BNL during the finishing stages of this work.
\end{acknowledgements}

%%%%%%%%%%%%%%%%%%%%%%%%%%%%%%%%%%%%%%%%%%%%%%%%%%%%%%%%%%%%%%%%%%%%%%%%%%%%%%
\bibliography{referencesnsi}

\end{document}